\documentstyle[aps,pra,epsfig,twocolumn]{revtex}

\def\be{\begin{equation}}
\def\ee{\end{equation}}
\def\bea{\begin{eqnarray}}
\def\eea{\end{eqnarray}}
\def\bma{\begin{mathletters}}
\def\ema{\end{mathletters}}
\def\C{\hbox{$\mit I$\kern-.7em$\mit C$}}

\begin{document}
\draft

\title{Activating bound entanglement in multi--particle systems}

\author{W. D\"ur and J. I. Cirac}

\address{Institut f\"ur Theoretische Physik, Universit\"at Innsbruck,
A-6020 Innsbruck, Austria}

\date{\today}

\maketitle

\begin{abstract}
We analyze the existence of activable bound entangled states in multi--particle
systems. We first give a series of examples which illustrate some different ways
in which bound entangled states can be activated by letting some of the parties
to share maximally entangled states. Then, we derive necessary conditions for a
state to be distillable as well as to be activable. These conditions turn out to
be also sufficient for a certain family of multi--qubit states. We use these
results to explicitely to construct states displaying novel properties related to
bound entanglement and its activation.
\end{abstract}


\narrowtext


\section{Introduction}

The existence of bound entanglement (BE) \cite{Ho97} has been one of the most 
intriguing surprises in quantum information during the last few years. A state 
is BE if (despite of being entangled) one cannot distill \cite{Be96} maximally 
entangled states (MES) out of it by using local operations and classical 
communication (LOCC). From this definition it follows that with bound entangled 
states (BES) one cannot perform reliable teleportation \cite{Be93,Li98}, quantum 
communication \cite{Noisy}, etc, i.e. they seem not to be useful for quantum 
information purposes. However, it has been shown that under certain conditions 
BE can be ``activated'' \cite{Ho99}. That is, with the help of some other 
entangled states they enable to carry out certain tasks which could not be 
performed by using these other entangled states and LOCC alone.

The first examples of BE arose in the context of {\it two systems}. In 
particular, the Horodecki showed that a necessary condition for a state of two 
systems to be distillable is that the corresponding density operator had 
nonpositive partial transposition \cite{Ho98,Du99b}. Thus, all entangled states 
with a positive partial transposition \cite{Pe96,Ho96} cannot be distilled, and 
therefore they are examples of BES \cite{Br00}. Very recently, we showed that a 
different kind of BE can also appear in {\em multiparticle systems} 
\cite{Du99a}. In particular, we considered states of three systems $A$, $B$, and 
$C$, spatially separated, that cannot be prepared locally and that have the 
following properties: (i) if we would be allowed to join systems $A$ and $C$ in 
one place, the state could be prepared locally; (ii) if we would be allowed to 
join systems $B$ and $C$ in one place, the state could be prepared locally; 
(iii) even if we would be allowed to join systems $A$ and $B$ in one place, the 
state could not be prepared locally. The property (i) immediately implies that 
one cannot distill a MES between the systems $A$ and $B$ (or $B$ and $C$) by 
using LOCC, since even if we would allow nonlocal operations between $A$ and $C$ 
we would not be able to do it. Analogously, the property (ii) implies that one 
cannot distill a MES between the systems $A$ and $C$. Thus, no MES can be 
distilled between any of the systems. The property (iii), however, indicates 
that the state is entangled, since even if we would allow nonlocal operations 
between $A$ and $B$ we could not prepare it locally. Thus, despite the state is 
entangled, one cannot distill any MES out of it, and therefore it is a BES. 
Nevertheless, the state presented in \cite{Du99a} can be activated. This follows 
from the fact that if we allow $A$ and $B$ to share some maximally entangled 
states (which is equivalent to allow $A$ and $B$ to join), then one can indeed 
distill a MES shared between $A$, $B$ and $C$. Another example of activable BE 
has been given by Smolin in the context of unlockable entanglement for the case 
of four systems \cite{Sm00}. In that case, the state has the extra property that 
the entanglement between $A$ and $B$ can be activated using a single copy of the 
state only and by letting $C$ and $D$ share only one MES. These examples show 
that a new kind of BE can arise when we split some free entanglement among more 
than two parties. They also show that the possibility of activating BE is 
sometimes counterintuitive and may have some applications related to the process 
of secret sharing \cite{Go99}.

In this paper we construct multiparticle states which display novel properties
related to BE. They illustrate the richness of multiparticle as compared to
two--particle entangled states regarding the possibility of activating BE. This
paper is organized as follows. In Sec.\ \ref{Activation}, we state the problem
of activation of BE in multiparticle systems and present various intriguing
examples of activable BE. The rest of the paper provides the tools needed to
construct states having the properties given in all those examples, as well as
the guidelines to construct other states with different properties. In Sec.\
\ref{splittings}, we review both the concept of bipartite splittings
\cite{Du99c} and the family of $N$ qubit states introduced in Ref.\
\cite{Du99a}. These states will play a central role throughout the remaining of
the paper; as we will show, they give rise to a vast variety of activable BES.
In Sec.\ \ref{bipartite}, we consider the distillability properties of a
multiparticle state when we allow the particles to join into two groups (i.e.
according to a bipartite splitting of the particles). We will show that it is
possible to find BES that can be activated iff the particles are joint according
to certain bipartite splittings. Moreover, we will show that one can choose the
bipartite splittings for which the BE can be activated without any restriction,
and that our family of states covers all possible examples of this kind. In
Sec.\ \ref{Necessary}, we consider a more general framework where the particles
are allowed to join into more than two groups. We derive necessary conditions
for distillability and activation based on the distillability properties of
bipartite splittings. We show that these conditions are also sufficient for
our family of states, which implies that they are an important ingredient to
construct generic examples of activable BES. In particular, they allow us to
construct BES fulfilling the properties introduced in the examples of Sec.\
\ref{Activation}. We finally summarize our results in Sec.\ \ref{Summary}.

\section{Activation of bound entanglement} \label{Activation}

Let us consider $N$ parties, $A_1,\ldots,A_N$, at different locations, each of
them possessing several qubits. We will assume that the state of the qubits is
described by a density operator of the form $\rho^{\otimes M}$, where $\rho$
denotes an entangled state of exactly $N$ qubits, each of them belonging to a
different party. Thus, the parties have $M$ copies of the same state, where $M$
can be as large as we wish. This ensures that the parties can use distillation
protocoles \cite{Be96} in order to obtain MES between some of them. In that case
we will say that the state $\rho$ is distillable (with respect to the specific
parties that can obtain a MES). A state $\rho$ is BE if it is not distillable
when the parties remain separated from each other. We say that a BES can be
activated if it becomes distillable once some of the parties join and form
groups to act together. Note that instead of allowing some parties to join we
could have allowed them to share some MES. In that case we would have the same
situation given the fact that separated parties sharing MES can perform any
arbitrary joint operation by simply teleporting back and forth the states of
their particles.


\subsection{Examples}

In this subsection we introduce some relevant examples of BES, by specifying 
their properties with respect to activation. In the following sections we will 
explicitely construct density operators that fulfill the properties given here 
for each of the examples. The goal of this subsection is to display the 
underlaying properties of activable BES.

\begin{figure}[ht]
\begin{picture}(230,200)
\put(0,55){\epsfxsize=230pt\epsffile[-1 705 202 830]{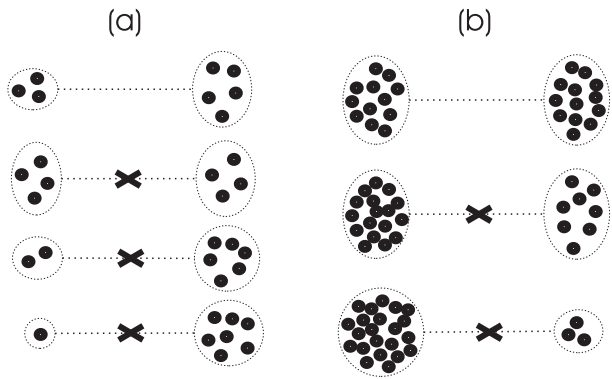}}
\end{picture}
\caption[]{In (a), we have $N=8$ and the state $\rho_I$ is distillable
iff 3 and 5 parties form a group, and $\rho_I$ is not distillable
otherwise. In (b), we have $N=25$ and $\rho_{II}$ is distillable iff two
groups of approximately the same size are formed (here 12-13), while
$\rho_{II}$ is not distillable iff one group contains less than $40\%$
of the particles $(j < 10)$. In both examples, it does not matter which
of the parties belongs to which group.}
\label{Fig1}
\end{figure}

\noindent
{\bf Example I:} The state $\rho_I$ becomes distillable iff the parties form two
groups with exactly $j$ and $N-j$ members, respectively. Furthermore, it does
not matter which of the parties join in each group, but only the number of
members. For example, if $N=8$ and $j=3$ (see Fig. \ref{Fig1}a), we have that
$\rho_I$ is distillable if exactly $3$ and $5$ parties join, but remains
undistillable when the parties form two groups with 1-7, 2-6, 4-4 members, or if
they form more than two groups. In particular, $\rho_{I}$ is not distillable if
the parties remain separated from each other, which corresponds to having $8$
groups.

\noindent
{\bf Example II:} The state $\rho_{II}$ is distillable iff the parties form two
groups, each of them containing say between $40\%$ and $60\%$ of the parties
(and where each party is contained in one of the two groups). In particular, the
state $\rho_{II}$ remains undistillable if the parties form more than two groups
or if they form two groups, with one of them containing less than $40 \%$ of the
parties. Again, it does not matter whether certain parties belong to the same
group; only the total number of particles within each group is important. We can
view $\rho_{II}$ as a state having ``macroscopic'' entanglement, but no
(distillable) ``microscopic'' entanglement (see Fig. \ref{Fig1}b). Note that one
can also have the opposite effect, i.e a state $\rho_{II}'$ which is distillable
iff the parties form two groups, and one of the groups contains less than say
$20\%$ of the parties.

\noindent
{\bf Example III:} The state $\rho_{III}$ becomes distillable iff the parties
form two groups, where the first group contains a {\it specific} set of $M$
parties $A=\{A_{k_1},\ldots A_{k_M}\}$, and the second group contains the
remaining parties. For all other configurations in groups $\rho_{III}$ remains
undistillable. For example, we have for $N=5$ and $A=\{A_1,A_3,A_5\}$ that
$\rho_{III}$ is distillable iff the the parties form two groups,
$(A_1A_3A_5)-(A_2A_4)$, and not distillable otherwise.

\begin{figure}[ht]
\begin{picture}(230,200)
\put(0,55){\epsfxsize=230pt\epsffile[1 716 190 838]{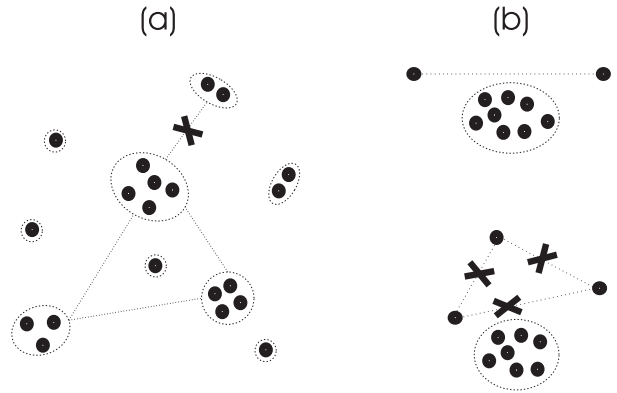}}
\end{picture}
\caption[]{In (a), we have that $N=20, j=3$ and $\rho_{IV}$ is such that
entanglement between any two groups which contain more or equal than $j=3$
members can be distilled (those groups can thus share a GHZ state), while groups
with less than 3 members can not create entanglement with any other group. In
(b), we have that $N=10$ and $\rho_{V}$ is such that once any $(N-2)$ parties
join, the remaining two can create entanglement, while no entanglement can be
shared among the remaining parties if less than $(N-2)$ parties join.}
\label{Fig2}
\end{figure}

\noindent
{\bf Example IV:} The state $\rho_{IV}$ has the following properties: Given any
two groups, each of them with a specific number $j$ (or more) parties, they can
distill (with the help of the other parties) a MES between them. In particular,
if there are several groups of $j$ or more parties, they can distill a GHZ--like
state \cite{Gr89} between all the groups. When a group with less than $j$
members is formed, it cannot distill a MES with any other group (see Fig.
\ref{Fig2}b). We call this effect clustering, since the parties have to form
clusters with at least $j$ members in order to be able to create MES. In a
similar way, one can also choose the state such that not only the number but
also the specific parties which have to form groups in order to create a MES is
given.

In the preceeding examples we have that some number of parties have to join into
groups in order to distill some MES with some other parties. In the following,
we will give examples in which the parties that join enable the remaining ones
to distill entanglement.

\noindent
{\bf Example V:} The state $\rho_{V}$ is such that once any $(N-2)$ parties form
a group, the remaining two parties can distill a MES, but $\rho_{V}$ remains
undistillable if less than $(N-2)$ parties join (see Fig. \ref{Fig2}b).

\begin{figure}[ht]
\begin{picture}(230,200)
\put(0,55){\epsfxsize=230pt\epsffile[27 683 233 824]{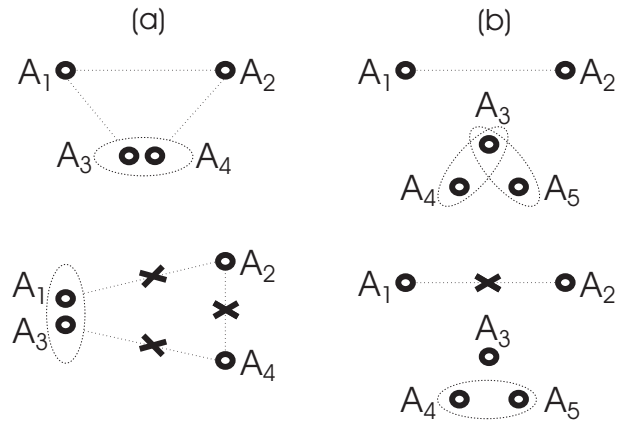}}
\end{picture}
\caption[]{In (a), we have $N=4$ and $\rho_{VI}$ is such that a GHZ state shared
among the groups $A_1-A_2-(A_3A_4)$ can be created iff the parties $(A_3A_4)$ join, and
no entanglement can be distilled if any other two parties form a group. In (b)
we have that $N=5$ and $\rho_{VII}$ is such that entanglement between $A_1$ and
$A_2$ can be created iff either the parties $(A_3A_4)$ or $(A_3A_5)$ join,
however no entanglement can be distilled if $(A_4A_5)$ join.}
\label{Fig3}
\end{figure}

\noindent
{\bf Example VI:} $\rho_{VI}$ is a state of $N=4$ parties for which once the
parties $(A_3A_4)$ form a group a GHZ--like state can be distilled among $A_1$,
$A_2$, and the group $(A_3A_4)$, whereas it is undistillable whenever any other
parties but $(A_3A_4)$ are joint (see Fig. \ref{Fig3}a). In contrast to the
previous example, in this one it is not only possible to create a MES between
$A_1$ and $A_2$ by joining $(A_3A_4)$, but also this last group can distill a
MES with the remaining parties.

\noindent
{\bf Example VII:} $\rho_{VII}$ is a state of $N=5$ parties such that a MES
between parties $A_1$ and $A_2$ can be created iff either the parties $(A_3A_4)$
or $(A_3A_5)$ join (see Fig. \ref{Fig3}b), but no entanglement can be distilled
if the parties $(A_4A_5)$ join.

In the following, we will explain how all those examples can be constructed and
understood by giving necessary conditions for distillation and activation in
multiparticle systems. These conditions also provide us with the tools to
construct other examples of activable BES. We will also introduce a family of
states which includes all the examples I-VII as well as all those examples which
can be constructed using the rules following from the necessary conditions for
distillation and activation, which are also sufficient for this family.

\section{Definitions and notation}\label{splittings}

In this section we first review the concept of bipartite splittings, and 
introduce some notation that will be used in the following sections. Then we 
review the properties of the family of $N$--qubit states $\rho_N$ introduced in 
Ref.\ \cite{Du99a}. As mentioned above, these states give rise to a wide range 
of activable BES. In particular, among them one can find states corresponding to 
the examples of activable BES introduced in the previous section.

\subsection{Bipartite splittings}

Let us denote by ${\cal P}$ the set of all possible (bipartite) splittings of 
$N$ parties into two groups. For example, for $3$ parties ${\cal P}$ contains 
the splittings $(A_1A_3)$--$(A_2)$, $(A_2A_3)$--$(A_1)$, and 
$(A_3)$--$(A_1A_2)$. We will denote these bipartite splittings by $P_{k}$, where 
$k=k_1k_2\ldots k_{N-1}$ is a chain of $N-1$ bits, such that $k_n=0,1$ if the 
$n$--th party belongs to the same group as the last party or not. For example, 
for $3$ parties the splittings $(A_1A_3)$--$(A_2)$, $(A_2A_3)$--$(A_1)$, and 
$(A_3)$--$(A_1A_2)$ will be denoted by $P_{01}$, $P_{10}$, and $P_{11}$, 
respectively. We will denote by $A$ the side of the splitting to which the party 
$N$ belongs and by $B$ the other side. In general, there exist $s=2^{N-1}-1$ of 
such splittings. In the following, when we consider bipartite splittings the 
parties in each of the groups will be allowed to act together (i.e. to perform 
joint operations).

\subsection{Family of states $\rho_N$}

Let us consider $\rho_N$, the family of $N$--qubit states introduced in
\cite{Du99a}. We have that $\rho\in\rho_N$ if it can be written as
\bea
\label{rhoN}
\rho &=& \sum_{\sigma=\pm} \lambda_0^\sigma |\Psi^\sigma_0\rangle\langle
  \Psi^\sigma_0| \nonumber\\
&& + \sum_{k\not= 0} \lambda_k (|\Psi^+_k\rangle\langle \Psi^+_k|
  + |\Psi^-_k\rangle\langle \Psi^-_k|),
\eea
where
\be
|\Psi^\pm_k\rangle \equiv \frac{1}{\sqrt{2}} (|k_1k_2\ldots k_{N-1} 0\rangle \pm
|{\bar k_1}{\bar k_2}\ldots {\bar k_{N-1}}1\rangle),
\ee
are GHZ--like states with $k=k_1k_2\ldots k_{N-1}$ being a chain of $N-1$ bits,
and ${\bar k_i}=0,1$ if $k_i=1,0$, respectively. We have that $\rho_N$ is
parametrized by $2^{N-1}$ independent real numbers. The labeling is chosen such
that $\Delta\equiv \lambda_0^+- \lambda^-_0 \ge 0$. As we will see below, both
the separability and distillability properties of the states belonging to this
family are completely determined by the coefficients
\be
s_{k}\equiv \left\{\begin{array}{l}
\mbox{1 if $\lambda_k < \Delta/2$} \\
\mbox{0 if $\lambda_k\geq \Delta/2$.}
\end{array}
\right.
\ee

Let us emphasize that the notation used for the states of this family parallels 
the one used to denote the partitions $P_k$. In fact, as shown in \cite{Du99c} 
the separability properties of $\rho_N$ for a given partition $P_k$ are directly 
related to the coefficient $s_k$:

{\bf Lemma 0:} \cite{Du99c} For any bipartite splitting $P_k \in {\cal P}$, and 
$\rho \in \rho_N$ we have $\rho^{T_{A}} \geq 0 \Leftrightarrow s_k=0 
\Leftrightarrow \rho$ is separable with respect to this 
splitting\footnote{$\rho^{T_{A}}$ denotes the partial transposition with respect 
to the parties $A$. For the definition of partial transpostion in mulitparticle 
systems see \cite{Pe96,Du99c}}.

Thus the coefficient $s_k$ determines whether $\rho$ is separable or not with 
respect the bipartite splitting $P_k$. Note that there are no restrictions to 
the values of these coefficients; that is, for any choice of $\{s_k\}$ there 
always exists a state $\rho\in\rho_N$ with these values. This automatically 
implies that the family $\rho_N$ provides us with all possible examples of 
states in which the separability properties of the bipartite splittings are 
completely specified. As we will see in the next section, the same is true 
regarding distillability with respect to bipartite splittings.

\section{Activation of BE for bipartite splittings: examples I-III}\label{bipartite}

With the states introduced in the previous section, we are at the position of 
examining the examples I--III. If we take a closer look at them, we find that 
they all have a common feature: the states $\rho_{I}$, $\rho_{II}$ and 
$\rho_{III}$ are distillable with respect to certain bipartite splittings, and 
not distillable with respect to other bipartite splittings. In this Section we 
will show that given a subset of all possible bipartite splittings we can always 
find states in $\rho_N$ such that they are distillable iff the parties join 
according to a bipartite splitting contained in this set. This general result 
clearly allows to find states within our family which correspond to examples 
I--III.

Let us be more specific. We consider the set of all possible bipartite 
splittings ${\cal P}$. Let us specify for each splitting $P_k$ whether we want 
that one can distill a MES or not. To do that, we will assign to each splitting 
a $0$ if we do not want it to be possible and a $1$ otherwise. That is, each 
possible specification corresponds to a function $f:{\cal P} \to\{0,1\}$. There 
are $2^s$ of such functions, which we will call specifications. Now, given a 
specification $f$ we define the set of splittings ${\cal S}_f=\{P\in {\cal P} 
{\mbox{ such that }}f(P)=1\}$. Thus, the problem reduces to finding a state 
$\rho$ such that only for the splitings contained in ${\cal S}_f$ one can 
distill a maximally entangled state. Note that examples I--III are just 
different specifications $f$. What we will show here is that there exist states 
$\rho\in\rho_N$ fulfilling any given specification.

\subsection{Family of states $\rho_N$}

We will show here that for the family of states $\rho_N$, bipartite 
inseparability is equivalent to distillability. This equivalence is expressed in 
the following Lemma:

{\bf Lemma 1:} Given the bipartite splitting $P_k\in {\cal P}$ and $\rho \in 
\rho_N$, we have that $\rho$ is distillable with respect to $P_k$ 
$\Leftrightarrow s_k=1$.

{\em Proof:} Using Lemma 0 we have that if $s_k=0$ then $\rho^{T_{A}}\ge 0$, and 
therefore the state is not distiallable, since non--positive partial 
transposition is a necessary condition for distillation \cite{Ho97}. Thus, we 
just have to show that if $s_k=1$ then the state is distillable. We denote by 
$|0\rangle_{A,B}$ and $|1\rangle_{A,B}$ the states in which all qubits in side 
$A$ or $B$ are in state $0$ and $1$, respectively. We have that 
$|\Psi_0^\pm\rangle =\frac{1}{\sqrt{2}} 
(|0\rangle_{A}|0\rangle_{B}\pm|1\rangle_{A}|1\rangle_{B})$ and 
$|\Psi_k^\pm\rangle =\frac{1}{\sqrt{2}} 
(|0\rangle_{A}|1\rangle_{B}\pm|1\rangle_{A}|0\rangle_{B})$. By measuring the 
projectors $|0\rangle_{A,B}\langle 0|+|1\rangle_{A,B}\langle 1|$ in $A$ and $B$ 
respectively, we only get contributions from the states $|\Psi_0^\pm\rangle$ and 
$|\Psi_k^\pm\rangle$ and one obtains that the state after a successful 
measurement is
\bea
\rho &\propto& \lambda_0^+ |\Psi_0^+\rangle\langle \Psi_0^+| +
\lambda_0^- |\Psi_0^-\rangle\langle \Psi_0^-| \nonumber\\
&&+\lambda_k (|0,1\rangle\langle 0,1| + |1,0\rangle\langle 1,0|.
\eea
This state is known to be distillable for $s_k=1$ \cite{Be96}.
$\Box$

This Lemma tells us that for a given specification we just have to find a state 
in $\rho_N$ such that $s_{k}=f(P_{k})$ for all $k$. Since the values $s_{k}$ are 
not restricted in any way we have that for any specification we can find states 
$\rho\in \rho_N$ fulfilling it, and thus our family provides all different kinds 
of bipartite distillable and not--distillable states.

Let us now come back to the examples I--III. We can take as state $\rho_{I}$ 
[example I] one from the family $\rho_N$ which has $s_{k}=f(P_{k})=1$ iff the 
number of ones in $k$ is $j$ or $(N-1-j)$ and $s_{k}=f(P_{k})=0$ otherwise (this 
means that all bipartite splittings which contain exactly $j$ members in one 
group are distillable, and all others are separable). In the example II, we 
choose $\rho_{II} \in \rho_N$ such that $s_{k}=f(P_{k})=1$ iff $P_k$ has between 
$40\%$ and $60\%$ of the parties in $B$. In the example III we take $\rho_{III} 
\in \rho_N$ such that $s_k=f(P_k)=1$ only for one specific $P_k$. We also 
emphasize that in a similar way one may construct many other interesting 
examples.

\section{Activation of BE by external action: Examples IV-VII}\label{Necessary}

While the separability and distillability properties with respect to bipartite 
splittings of a state $\rho$ were sufficient to completely understand and 
construct examples I-III, in the remaining examples we are faced with a slightly 
more complicated situation. Now we have that the parties can join into more than 
two groups, and it may even happen that not all of the parties are needed in 
order to activate the BE. Thus, the bipartite splittings $P_k$ do no longer 
provide a complete description of the problem. However, we can still use them to 
derive {\it necessary} conditions regading distillability and even activation of 
BE between any number of groups. And what is more important, we will show that 
these necessary conditions turn out to be also {\it sufficient} for the family 
of states $\rho_N$. On the one hand, this allows us to construct various 
different kinds of activable BES, such as those corresponding to examples 
IV-VII. On the other hand, it ensures that the family $\rho_N$ provides examples 
for all possible kinds of BES which can be constructed by using the necessary 
conditions for distillation based on the bipartite properties of a state $\rho$. 
We wish to emphasize that due to the fact that the conditions we obtain are only 
necessary and not sufficient in general, there may exist other kind of activable 
BES that cannot be obtained with the methods developed here. Nevertheless, these 
methods also give indications about some other kinds of BES.


\subsection{Necessary conditions for distillation}

The separability properties of the bipartite splittings $P_k$ of a multiparticle 
state $\rho$ provide necessary conditions for the distillability properties of 
$\rho$. This is expressed in the following result:

{\bf Theorem 1:} Let $C=\{A_{i_1},\ldots,A_{i_M}\}$ and 
$D=\{A_{j_1},\ldots,A_{j_L}\}$ be two disjoint groups of $M$ and $L$ parties 
respectively, whereas the rest of the parties are separated. If a MES between 
$C$ and $D$ can be distilled then $\rho$ has to be non--separable with respect 
to all those bipartite splittings $P_k$ in which the groups $C$ and $D$ are 
located on different sides.

{\em Proof:} Let us assume that $\rho$ is separable with respect to one of those 
bipartite splittings $P_k$, so that $C\subset A$ and $D\subset B$. This means 
that even if we allow the groups $C$ and $D$ to join some other parties 
(belonging to $A$ and $B$, respectively), they will not be able to distill a 
MES. This is due to the fact that nonseparability is a necessary condition for 
distillability. $\Box$

Theorem 1 relates the distillability properties of $\rho$ to the classification 
with respect to the separability properties of a multiparticle state given in 
\cite{Du99c}. We also note here that the $k$--separability properties with 
respect to $k$--partite splittings ($k >2$) provide no additional information 
about the distillability properties of a state $\rho$. Theorem 1 also determines 
the necessary conditions for the creation of GHZ--like states, since the 
possibility of creating maximally entangled pairs between any two out of $l$ 
parties is a necessary and sufficient condition for the creation of a GHZ--like 
state among those $l$ parties. We can also change ''non--separable'' in the 
theorem to ''distillable'', which provides an even stronger condition. It is not 
clear whether this condition is then also sufficient. One may think of the 
existence of bound entangled states which are distillable with respect to all 
possible bipartite splittings, but which are not distillable when considering 
the parties independently. Recently, it has been reported that such states in 
fact exist \cite{Sm00b}.

We also see from Theorem 1 that in order for a MES between two specific, 
separated parties, to be distillable, it is necessary that the corresponding 
state is inseparable with respect to at least $2^{N-2}$ different bipartite 
splittings. Thus there exist many states which are inseparable, but do not 
fulfill the necessary condition for distillability between any two parties. All 
those states are obviously BE. For example, any state which has less than 
$2^{N-2}$ (and more than one) inseparable bipartite splittings is clearly bound 
entangled. Naturally, the question arises under which conditions this BE can be 
activated.

\subsection{Necessary conditions for activation of BES}

Clearly, the situation changes when the parties are allowed to form groups and 
act together. In this case, the parties may be able to change the separability 
properties of certain bipartite splittings $P_k$. For example, for $N=3$ parties 
$A, B$, and $C$, if the parties $A$ and $B$ form a group they may be able to 
change the separability properties of the bipartite splittings (i) $(B)$--$(AC)$ 
and (ii) $(A)$--$(BC)$. This is due to the fact that 'joining' is equivalent to 
having some extra entanglement available. In this case, this extra entanglement 
between $A$ and $B$ can be used to change the separability properties of the 
bipartite splittings in question. Note, however, that this extra entanglement 
does not help to change the separability properties of the bipartite splitting 
(iii) $(C)$--$(AB)$, since $(AB)$ where already joint in this bipartite 
splitting. This also allows us to understand the example of an activable BE 
three party state given in the introduction \cite{Du99a}, where we had that (i) 
and (ii) are separable, while (iii) is inseparable and the state is thus BE 
according to Theorem 1. By joining the parties $A$ and $B$, one may however 
change the bipartite splittings (i) and (ii) from separable to inseparable, so 
the necessary conditions for distillation may now be fulfilled, since all 
bipartite splittings can now be inseparable in principle. As shown in 
\cite{Du99a}, this activation can in fact be achieved, i.e. the change of the 
separability properties of the splittings (i) and (ii) as well as the 
distillation are possible. In a similar way, one can explain the example given 
in \cite{Sm00} for $N=4$. The effect of activation of BE for a state $\rho$ can 
thus be viewed as a consequence of the following theorem:

{\bf Theorem 2:} Consider a state $\rho$ which is separable with respect to a 
given bipartite splitting $(A)$--$(B)$. When joining $M$ parties 
$C=\{A_{i_1},\ldots,A_{i_M}\}$, a necessary condition that we can make $\rho$ 
distillable with respect to the splitting $(A)$--$(B)$ is that: (i) 
$C\not\subset A,B$; (ii) by using an operation acting on $C$ one can transform 
the state such that it is now nonseparable with respect to the bipartite 
splitting $(A)$--$(B)$.

{\em Proof:} (ii) follows trivially from Theorem 1, whereas (i) follows from 
(ii). $\Box$

According to this theorem, when joining some parties into a group $C$, they may 
change the separability and distillability properties of certain bipartite 
splittings, namely all those splittings $(A)$--$(B)$ for which $C\not\subset 
A,B$. So, it may happen that the conditions for distillability, which were not 
fulfilled before joining the parties, are now fulfilled, and a MES shared 
between some parties can now be created in principle. Theorems 1 and 2 together 
provide necessary conditions for the activation of multiparticle BES and provide 
thus the framework for the construction of generic examples of different kinds 
of activable bound entangled states. To achieve this, one chooses at the 
beginning the separability and distillability of the bipartite splittings $P_k$ 
of a state $\rho$ such that the state is not distillable if the parties remain 
separated from each other, but that the distillability conditions may be 
fulfilled when some of the parties are allowed to form groups. We also note here 
that due to the fact that both theorems only provide necessary conditions in 
general, we do not have that the distillabity and activation properties of a 
state $\rho$ are completely determined by the separability properties of its 
bipartite splittings. On the one hand, there might exist states which cannot be 
distilled or activated eventhough they fulfill the necessary conditions for 
distillation/activation, i.e. those states are further protected against 
activation. On the other hand, we already see that there are various kinds of 
non--activable bound entangled states. For example, all states which are 
biseparable with respect to all bipartite splittings but are inseparable with 
respect to any $k$--partite splitting ($k>2$) (see also classification proposed 
in \cite{Du99c}) are clearly bound entangled and can be neither distilled nor 
activated. For $N=3$, such an example is known \cite{Be98}.

Using Theorems 1 and 2, it is now straightforward to check that in example I the 
state $\rho_I$ reamains undistillable whenever the parties form more than two 
groups, since the necessary condition for distillation of a MES between any two 
groups can not be fulfilled. On the other hand, it is already clear that if the 
parties form two groups with a different number of members than $j$ and $N-j$, 
we have by construction that the state $\rho_{I} \in \rho_N$ is separable with 
respect to this bipartite splitting and thus not distillable. In a similar way, 
one can check that in examples II and III the states $\rho_{II}$ and 
$\rho_{III}$ are only distillable iff the parties join in two groups of required 
size (example II) or required members (example III), respectively.

\subsection{Family of states $\rho_N$}

In this section, we show that the necessary condition for distillation and 
activation expressed in Theorems 1 and 2 are also sufficient for the family of 
states $\rho_N$. We first show that if the necessary conditions for the 
distillation of MES between any two groups of parties, both not including a 
certain party $A_l$, are fulfilled, one can disentangle party $A_l$ from the 
remaining system while keeping the necessary conditions for distillation between 
the two groups in question. In order to achieve this, the party $A_l$ has to 
cooperate, i.e. its help is requried. This is expressed in the following lemma:

{\bf Lemma 2:} One can convert any $N$--qubit state $\rho \in \rho_N$ to a 
$(N-1)$--qubit state $\tilde\rho \in \rho_{N-1}$ by measuring a certain party 
$A_l$, such that for all bipartie splittings $P_k$ the following property is 
fulfilled: If $\rho$ is inseparable with respect to the bipartite splittings 
$(A_l C )$--(rest) and $(C )$--$(A_l$ rest) $\Rightarrow \tilde\rho$ is 
inseparable with respect to the bipartite splitting $(C)$--(rest).

{\em Proof:} We assume without loss of generality that $A_l=A_1$. If we measure 
in $A_1$ the Projector $P_+=|+\rangle_{A_1}\langle+|$ where 
$|+\rangle=(|0\rangle+|1\rangle)/\sqrt{2}$, we find that the remaining $(N-1)$ 
particles are in a state of the form $\rho_{N-1}$ with new coefficients 
$\tilde\lambda_{j_2j_3\ldots j_{N-1}}=\lambda_{0j_2j_3\ldots 
j_{N-1}}+\lambda_{1j_2j_3\ldots j_{N-1}}$ and similar for $\lambda_0^\pm$, so 
that $\tilde\Delta=\Delta$. The property we want to show is simply that iff both
\be
\lambda_{0j_2j_3\ldots j_{N-1}}<\Delta/2 \mbox{ and }
\lambda_{1j_2j_3\ldots j_{N-1}}<\Delta/2 \label{*},
\ee
it follows that
\be
\tilde\lambda_{j_2j_3\ldots j_{N-1}}<\tilde\Delta/2 \label{**}.
\ee
It may happen that although (\ref{*}) is fulfilled, (\ref{**}) is not. In this 
case, we apply the first step of the distillation procedure proposed in 
\cite{Du99c} first, where one measures certain POVM elements on $M$ copies of 
the initial state and is left with a new (unnormalized) state of the form 
$\rho_N$ with new coefficients $\lambda'_k=\lambda_k^M$ and 
$(\Delta'/2)=(\Delta/2)^M$. For $M$ sufficiently large, we can now allways have 
that after applying $P_+$, the remaining state is such that (\ref{**}) is 
fulfilled, since the new $\Delta/2$ is exponentially amplified compared to any 
$\lambda_{k_{1,2}}<\Delta/2$ and thus 
$(\Delta/2)^M>\lambda_{k_1}^M+\lambda_{k_2}^M$ for $M$ sufficiently large. 
$\Box$

{\bf Theorem 3:}  For the family of states $\rho_N$, we have that the necessary
condition for distillability given in Theorem 1 is also sufficient.

{\em Proof:} To show this, one just has to sequentially apply Lemma 2 to all 
particles $A_i \notin \{C ,D\}$, which leaves us with a $(M+L)$ qubit--state 
$\rho \in \rho_{M+L}$ which has $\rho^{T_{C }} \not\geq 0$ (which, according to 
Lemma 0 means that the corresponding $s_k=1$) and thus Lemma 1 can be applied, 
ensuring that any state $\rho\in\rho_N$ fullfilling the necessary conditions for 
any kind of distillation is in fact distillable. $\Box$

Note that the help of {\it all} parties is required, independent of whether they 
finally share a MES or not. This follows from the fact that Lemma 2 is based on 
the cooperation of party $A_l$, which is separated from the remaining system 
after the procedure described above. If one would just trace out party $A_l$ 
(i.e. the party does not cooperate with the remaining parties), one finds that 
the remaining parties cannot create any entanglement at all.

{\bf Theorem 4:} For $\rho \in \rho_N$, and the situation of Theorem 2, we can 
in fact change {\it all} those bipartite splittings $(A)$-$(B)$ for which 
$C\not\subset A,B$ from separable to inseparable (which is equivalent to 
distillable in this case) without changing the separability properties of the 
remaining bipartite splittings.

{\em Proof:} We assume without loss of generality that we join the first $M$ 
parties $C =\{A_1,\ldots,A_M\}$. Let $j=j_1j_2\ldots j_M$ $(l=l_1\ldots 
l_{N-M-1})$ be $M$ [$(N-M-1)$] digit binary numbers. We have to show that we can 
change all those bipartite splittings which do not contain all parties $C$ on 
one side from separable to inseparable. This is equivalent to showing that for 
all $\lambda_{jl}$ with $j \not= \{0,2^M-1\}$, we can have $\Delta/2 > 
\lambda_{jl}$. When applying the POVM element $\tilde P=\sum_{j=0}^{2^M-1} 
\sqrt{y_j}|j\rangle_{C }\langle j|$ , we find that we obtain again an 
(unnormalized) state of the form $\rho_N$ with new coefficients 
$\tilde\lambda_{jl}=y_{j}\lambda_{jl}$. Choosing $y_0=y_{2^M-1}=1$ and all other 
$y_j$ sufficiently small, we have that $\tilde\Delta=\Delta$ and for $j \not= 
\{0,2^M-1\}$ we can obtain that $\tilde\Delta/2 > \tilde\lambda_{jl}$ as 
required. Furthermore, all other relations $\Delta/2 \times \lambda_{jl}$, with 
$\times \in \{>,\leq\}$ and thus the separability properties of all those 
bipartite splittings for which $C\subset A,B$ remain unchanged. $\Box$

Theorems 3 and 4 together ensure that any BES within the family $\rho_N$ which 
is activable in principle can in fact be activated, i.e. the necessary 
conditions for the activation of bound entangled states given in Theorem 1 and 2 
are also sufficient for the family $\rho_N$.


\subsection{Examples IV-VII}

We are now at the position to construct and explain examples IV--VII, as well as 
to provide many other interesting examples of activable BES.

Let us start with example IV: In this case, we choose the state $\rho_{IV} \in 
\rho_N$ such that it is separable with respect to all bipartite splittings $P_k$ 
where either the group $A$ or $B$ has less than $j$ members. All other bipartite 
splittings are chosen to be distillable. It is clear that if a group with less 
than $j$ members is formed, they cannot distill a MES with any other group 
(since the corresponding bipartite splitting is separable). However once the 
parties form two groups, each having more than $j$ members, it is 
straightforward to check (using Theorems 4 and 3) that these two groups can in 
fact create a MES.

In example V, we choose the state $\rho_V \in \rho_N$ such that it is 
distillable with respect to all bipartite splittings $P_k$ where either the 
group $A$ or $B$ contains exactly one particle (so that the number of ones in 
$k$ is either 1 or $N-1$), and separable with respect to all other bipartite 
splittings $P_k$. It is clear that it is sufficient to join any $N-2$ particles, 
since this allows - according to Theorem 4 - to make inseparable (and thus 
distillable) all those bipartite splittings where the remaining particles are 
located in different groups, so that distillation of a MES between the remaining 
two particles becomes possible, according to Theorem 3. On the other hand, if 
less than $N-2$ parties join, one can easily verify that for any two of the 
remaining parties, there always remains at least one bipartite splitting 
separable which has to be inseparable in order that distillation can be 
possible. Thus the remaining parties cannot share entanglement if less than 
$N-2$ parties join.

In example VI we choose $\rho_{VI} \in \rho_4$ such that it is inseparable with 
respect to the bipartie splittings $(A_1A_2)$--$(A_3A_4)$, 
$(A_1)$--$(A_2A_3A_4)$ and $(A_2)$--$(A_1A_3A_4)$ and separable with respect to 
all other bipartite splittings. Clearly, this state is BE since at least 
$2^{N-2}=2^2=4$ bipartite splittings have to be inseparable so that a pair 
between any two separated parties can be distilled. Furthermore, one can check 
(using Theorems 3 and 4) that $\rho_{VI}$ remains undistillable when joining any 
two parties but $(A_3A_4)$, but one can create a GHZ state shared among 
$A_1$--$A_2$--$(A_3A_4)$ once one joins the parties $(A_3A_4)$.

Finally, in example VII we have $N=5$ and choose $\rho_{VII} \in \rho_5$ such 
that the state is inseparable with respect to all bipartite splittings which 
contain $A_1$ and $A_2$ in different groups, except the splitting 
$(A_1A_3)$--$(A_2A_4A_5)$ which is chosen to be separable as well as all other 
bipartite splittings. One can readily check that $\rho_{VII}$ has the desired 
properties, i.e. it is BE and can be activated when joining the parties 
$(A_3A_4)$ or $(A_3A_5)$.

If we demand however that entanglement between $A_1$ and $A_2$ should be created 
once {\it any} two of the remaining parties join, one finds that such a a state 
cannot be constructed using our method. In this case, we have to demand that the 
initial state $\rho_{VII} \in \rho_N$ has to be separable with respect to at 
least one of the bipartite splittings where $A_1$ and $A_2$ belong to different 
groups. Otherwise, a MES between $A_1$ and $A_2$ could be distilled from the 
beginning and the state is thus not BE. Let us assume without loss of generality 
that the separable splitting is either (i) $(A_1A_3)$--$(A_2A_4A_5)$ or (ii) 
$(A_1)$--$(A_2A_3A_4A_5)$. In both cases, the separability properties of this 
splitting cannot be changed when joining the parties $(A_4A_5)$ and so no 
entanglement can be created between $A_1$ and $A_2$ although two of the 
remaining parties join (note that if we had taken any other bipartite splitting 
with $A_1$ and $A_2$ belonging to different groups to be separable, we would 
have found some other two of the remaining parties which can join but not change 
the properties of this splitting).

Due to the fact that the conditions for distillation and activation we give here 
are only necessary in general, they do not allow us to rule out the possibility 
that a state having the desired properties can exist. In this case, the 
activation would not be based on the change of the separability properties of 
the bipartite splittings, but on some other mechanism.

\section{Summary}\label{Summary}

In summary, we have given rules to construct activable bound entangled states
using the separability and distillablity properties of a density operator with
respect to bipartite splittings. This method allows us to construct examples for
all possible kinds of activable BES where the parties join into exactly two
groups. In particular, the family of states introduced in Ref.\ \cite{Du99a}
contains examples of all these kinds of activable BES. We have also given some
relevant examples of activation of BE in which the parties join into more than
two groups, and where the role of some of the groups is just to help the others
to create a MES.

We thank G. Vidal and J. Smolin for discussions. This work was supported by the 
Austrian Science Foundation under the SFB ``control and measurement of coherent 
quantum systems´´ (Project 11), the European Community under the TMR network 
ERB--FMRX--CT96--0087, the European Science Foundation and the Institute for 
Quantum Information GmbH.



\end{document}